\documentclass[aps,prl,article,twocolumn,citeautoscript,superscriptaddress]{revtex4-2} \synctex=1
\usepackage{amsmath,amssymb,bm,bbm}
\usepackage{amsthm}
\usepackage{physics}
\usepackage{graphicx}
\usepackage{color}
\usepackage[dvipsnames]{xcolor}
\usepackage[colorlinks=true]{hyperref}
\usepackage[section]{placeins}
\usepackage[mathscr]{euscript}
\usepackage[toc,page]{appendix}
\hypersetup{
    unicode=false,          
    pdftoolbar=true,        
    pdfmenubar=true,        
    pdffitwindow=false,     
    pdfstartview={FitH},    
    pdftitle={},    
    pdfauthor={},     
    pdfsubject={},   
    pdfcreator={},   
    pdfproducer={}, 
    pdfkeywords={} {} {}, 
    pdfnewwindow=true,      
    colorlinks=true,       
    linkcolor=magenta, 
    citecolor=blue,        
    filecolor=magenta,      
    urlcolor=blue           
}

\usepackage{pdfpages} 
\usepackage{pgffor} 

\makeatletter
\AtBeginDocument{\let\LS@rot\@undefined}
\makeatother

\def\supplementfilename{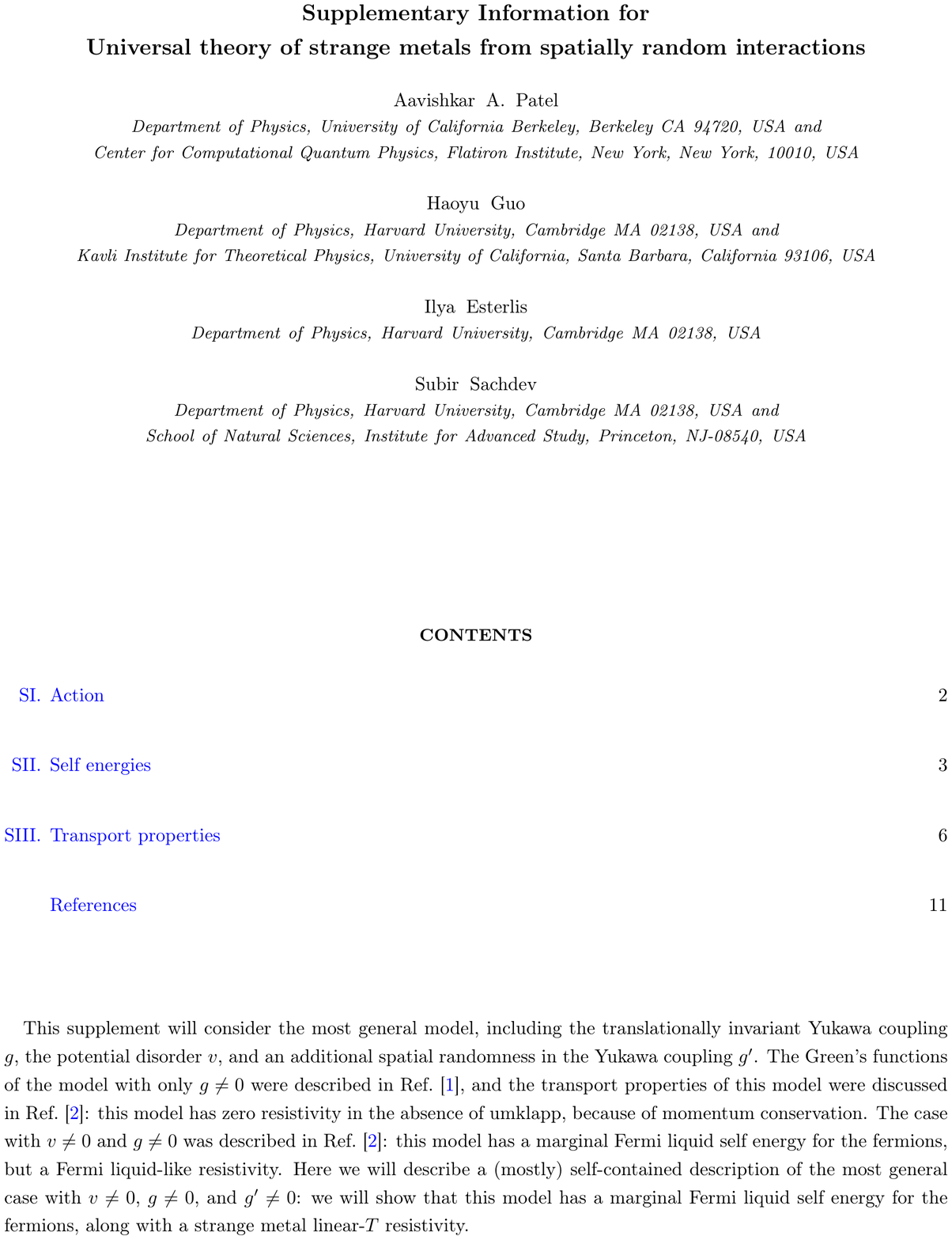}

\pdfximage{\supplementfilename}
\def\numbersupplementpages{\the\pdflastximagepages}

\newif\ifarXiv
\arXivtrue

\usepackage{amsfonts}
\usepackage{upgreek}
\usepackage{slashed}
\usepackage{latexsym}
\usepackage{mathrsfs}
\usepackage[usestackEOL]{stackengine}
\newcommand{\bk}{{\bm k}}
\newcommand{\bq}{{\bm q}}
\newcommand{\br}{{\bm r}}
\newcommand{\nn}{\nonumber \\}
\newcommand{\beq}{\begin{equation}}
\newcommand{\eeq}{\end{equation}}
\def\bea{\begin{eqnarray}}
\def\eea{\end{eqnarray}}
\allowdisplaybreaks

\begin{document}
\title{Universal theory of strange metals from spatially random interactions}

\author{Aavishkar A. Patel}
\affiliation{Center for Computational Quantum Physics, Flatiron Institute, New York,
New York, 10010, USA}
\affiliation{Department of Physics, University of California Berkeley, Berkeley CA 94720, USA}

\author{Haoyu Guo}
\affiliation{Laboratory of Atomic and Solid State Physics, Cornell University,
142 Sciences Drive, Ithaca NY 14853-2501, USA}
\affiliation{Department of Physics, Harvard University, Cambridge MA 02138, USA}
\affiliation{Kavli Institute for Theoretical Physics,
University of California, Santa Barbara, California 93106, USA}

\author{Ilya Esterlis}
\affiliation{Department of Physics, University of Wisconsin-Madison, Madison, Wisconsin 53706, USA}
\affiliation{Department of Physics, Harvard University, Cambridge MA 02138, USA}

\author{Subir Sachdev}
\affiliation{Department of Physics, Harvard University, Cambridge MA 02138, USA}
\affiliation{School of Natural Sciences, Institute for Advanced Study, Princeton, NJ-08540, USA}

\date{\today.~
\href{https://arxiv.org/abs/2203.04990}{arXiv:2203.04990}}

\begin{abstract}
We consider two-dimensional metals of fermions coupled to quantum critical scalars, the latter representing order parameters or fractionalized particles. We show that at low temperatures ($T$), such metals generically exhibit strange metal behavior with a $T$-linear resistivity arising from spatially random fluctuations in the fermion-scalar Yukawa couplings about a non-zero spatial average. We also find a $T\ln (1/T)$ specific heat, and a rationale for the Planckian bound on the transport scattering time. These results are obtained in the large $N$ expansion of an ensemble of critical metals.
\end{abstract}

\maketitle

A major theme in the study of correlated metals has been their strange metal behavior at low temperatures {\it i.e.\/} a linear-in-temperature resistivity smaller than the quantum unit of resistivity ($h/e^2$ in two dimensions) which appears to be controlled by a dissipative `Planckian' relaxation time of order $\hbar/(k_B T)$ (where $T$ is the absolute temperature) \cite{Hartnoll:2021ydi,Chowdhury:2021qpy,qptbook,bruin,Zaanen,Gael21,Paschen22,Sankar2022}. 
This behavior is in sharp constrast to $T^2$ dependence of the resistivity, and the $1/T^2$ relaxation time, invariably observed in conventional metals described by Fermi liquid theory.
Moreover, the anomalous resistivity of strange metals is accompanied by a logartihmic enhancement of the Sommerfeld metallic specific heat to $T \ln (1/T)$ \cite{Hartnoll:2021ydi} from the $\sim T$ behavior of conventional metals.

Starting with the seminal work of Hertz \cite{hertz}, there has been extensive research on the properties of electronic Fermi surfaces at quantum phase transitions \cite{SungSik18}. The quantum critical fluctuations are represented by a scalar field, which is usually a symmetry-breaking order parameter, but could also be a fractionalized particle at phase transitions without an order parameter \cite{SVS04}. This scalar field has a `Yukawa' coupling to the electrons, by which the electrons scatter by emitting or absorbing a scalar field excitation (the Yukawa coupling is similar to the electron-phonon coupling, but without a suppression by the gradient of the scalar field). It is now known that such a Fermi surface coupled to a quantum-critical scalar leads to a breakdown of the electronic quasiparticle excitations in two spatial dimensions \cite{PALee89,SungSik18}. But the Fermi surface survives as a sharp boundary in momentum space, separating particle- and hole-like excitations which are diffuse in energy space. In the presence of random impurities which scatter the electrons \cite{HLR,Rosch99,ChubukovMaslov,BergWang,criticalII}, there are cases where the quasiparticles are at the boundary of stability, leading to `marginal' Fermi liquid behavior \cite{MFL89} in single-particle observables such as those observed in photoemission experiments. 

However, despite these successes, theory has so far been unable to explain the defining transport properties (such as the linear-in-$T$ resistivity) of strange metals. 
Conservation of momentum in the low energy theory of a clean metal implies that the d.c. and optical conductivities are not affected by the anomalous self energy of the excitations near the Fermi surface \cite{Hartnoll:2007ih,ChubukovMaslov,Hartnoll:2014gba,Eberlein:2016jlt,BergWang,SenthilShi22,criticalII}. In other words, the strong coupling between the Fermi surface and the scalar field places the system in the limit of strong `scalar drag', and this clean theory cannot describe strange metal behavior. This is in contrast to the electron-phonon system, where the weak electron-phonon coupling makes phonon drag a factor only in ultrapure samples \cite{Hicks12}. Umklapp scattering can lead to non-zero resistance, and its influence in quantum-critical metals has been investigated in other works \cite{BergWang,Lee21}. But, umklapp is suppressed at low $T$, its predictions for transport are not universal and depend upon specific Fermi surface details, and there is no corresponding $T \ln (1/T)$ specific heat.

Given the ubiquity of strange metal transport across numerous correlated electron materials (from the cuprates and the pnictides to recently discovered twisted bilayer graphene), a simple and universal mechanism is called for. We propose here that spatial disorder in the fermion-scalar Yukawa coupling, about a non-zero spatial average, provides just such a mechanism. Such disorder is ubiquitous in correlated electron materials {\it e.g.\/} in a Hubbard model with on-site repulsion $U$, and an impurity-induced disorder in the electron hopping $t_{ij}$, the Schrieffer-Wolff transformation generates disorder in the exchange interaction $J_{ij} = 4 t_{ij}^2/U$, and this disorder then feeds into the Yukawa coupling after a standard decoupling procedure \cite{hertz} that introduces the scalar field. Moreover, our mechanism applies universally across different classes of quantum-critical metals, with scalars which are either fractionalized particles or order parameters at zero or non-zero momentum, which have distinct critical behaviors 
in the clean limit. We find a universal phenomenology that matches observations, including the $T$-linear resistivity, the Planckian relaxation time, and the $T \ln (1/T)$ specific heat.

A key observation of our analysis is that while the fermion inelastic self energy corrections can be dominated by the spatially uniform coupling, the transport is nevertheless dominated by the spatially random coupling, and this leads to our main results. Our work follows other recent works with random Yukawa interactions \cite{Fu:2016vas,Murugan:2017eto,Patel:2018zpy,Marcus:2018tsr,Wang:2019bpd,Ilya1,Wang:2020dtj,Altman1,WangMeng21,Esterlis:2021eth} inspired by the Sachdev-Ye-Kitaev (SYK) model \cite{SY92,kitaev2015talk},
along with studies which found linear-in-$T$ resistivity with random interactions, but with vanishing spatial average  \cite{Patel:2019qce,Guo:2020aog,Altman1,Dumi21,Esterlis:2021eth}.



{\it Spatially uniform quantum-critical metal.} We begin by recalling the SYK-inspired large $N$ theory of the two-dimensional quantum-critical metal \cite{Esterlis:2021eth,Altman1} for the case where the order parameter has zero momentum. The imaginary time ($\tau$) action for the fermion field $\psi_i$ and scalar field $\phi_i$ (with $i=1 \ldots N$ a flavor index) is \cite{Esterlis:2021eth}
\begin{align}
&\mathcal{S}_g = \int d\tau\sum_{\bk}\sum_{i=1}^N\psi^\dagger_{i\bk}(\tau)\left[\partial_\tau +\varepsilon (\bk) \right]\psi_{i\bk}(\tau) \nn
&+\frac{1}{2}\int d\tau \sum_{\bq}\sum_{i=1}^N \phi_{i\bq}(\tau)\left[-\partial_\tau^2 + K \bq^2 +m_b^2\right]\phi_{i,-\bq}(\tau) \nn
&+\frac{g_{ijl}}{N} \int d\tau d^2 r \sum_{i,j,l=1}^N \psi^\dagger_{i}(\br,\tau)\psi_{j}(\br,\tau)\phi_{l}(\br,\tau)\,,
\label{eq:latticeaction}
\end{align}
where the fermion dispersion $\varepsilon(\bk)$ determines the Fermi surface, the scalar mass $m_b$ has to be tuned to criticality and is needed for infrared regularization but does not appear in final results, and $g_{ijl}$ is space independent but random in flavor space with
\beq
\overline{g_{ijl}} = 0 \,, \quad  \overline{{g}^\ast_{ijl}g_{abc}} = {g}^2\,\delta_{ia}\delta_{jb}\delta_{lc}\,,
\eeq
where the overline represents average over flavor space.
The hypothesis is that a large domain of flavor couplings all flow to the same universal low energy theory (as in the SYK model), so we can safely examine the average of an ensemble of theories. Momentum is conserved in each member of the ensemble, and the flavor-space randomness does not lead to any essential difference from non-random theories. This is in contrast to position-space randomness which we consider later, which does relax momentum and modify physical properties.

The disorder average of the partition function of $\mathcal{S}_g$ leads to a `$G$-$\Sigma$' theory, whose large $N$ saddle point of (\ref{eq:latticeaction}) has singular fermion ($\Sigma$) and boson ($\Pi$) self energies at $T=0$  \cite{Esterlis:2021eth}
\begin{align}
    \Pi(i\omega, \bq) =
    -c_b \frac{|\omega|}{|\bq|}\,,& \quad \Sigma (i\omega, \bk) = - i c_f \mbox{sgn}(\omega) |\omega|^{2/3}  \nonumber \\
      c_b= \frac{g^2}{2 \pi \kappa v_F}\,,& \quad c_f = \frac{g^2}{2 \pi v_F \sqrt{3}} \left( \frac{2 \pi v_F \kappa}{K^2 g^2} \right)^{1/3}\,.
\label{eq:sigmag}     
\end{align}
These results are obtained on a circular Fermi surface with curvature $\kappa=1/m$ where $m$ is the effective mass of the fermions. The result is consistent with the theory of two antipodal patches around $\pm \bk_0$ on the Fermi surface to which $\bq$ is tangent, with axes chosen so that $\bq = (0,q)$ and fermionic
dispersion $\varepsilon(\pm \bk_0 + \bk) = \pm v_F k_x + \kappa k_y^2/2$.

The large $N$ computation of the conductivity \cite{criticalII,SenthilShi22} yields only the clean Drude result $\mathrm{Re}[\sigma(\omega)]/N=\pi\mathcal{N}v_F^2\delta(\omega)/2$, where $\mathcal{N}=m/(2\pi)$ is the fermion density of states at the Fermi level. This is in contrast to the claimed \cite{PALee89,Kim94} d.c. resistivity $\sim T^{4/3}$ and optical conductivity  $\sim |\omega|^{-2/3}$.

\begin{figure}
\includegraphics[width=0.48\textwidth]{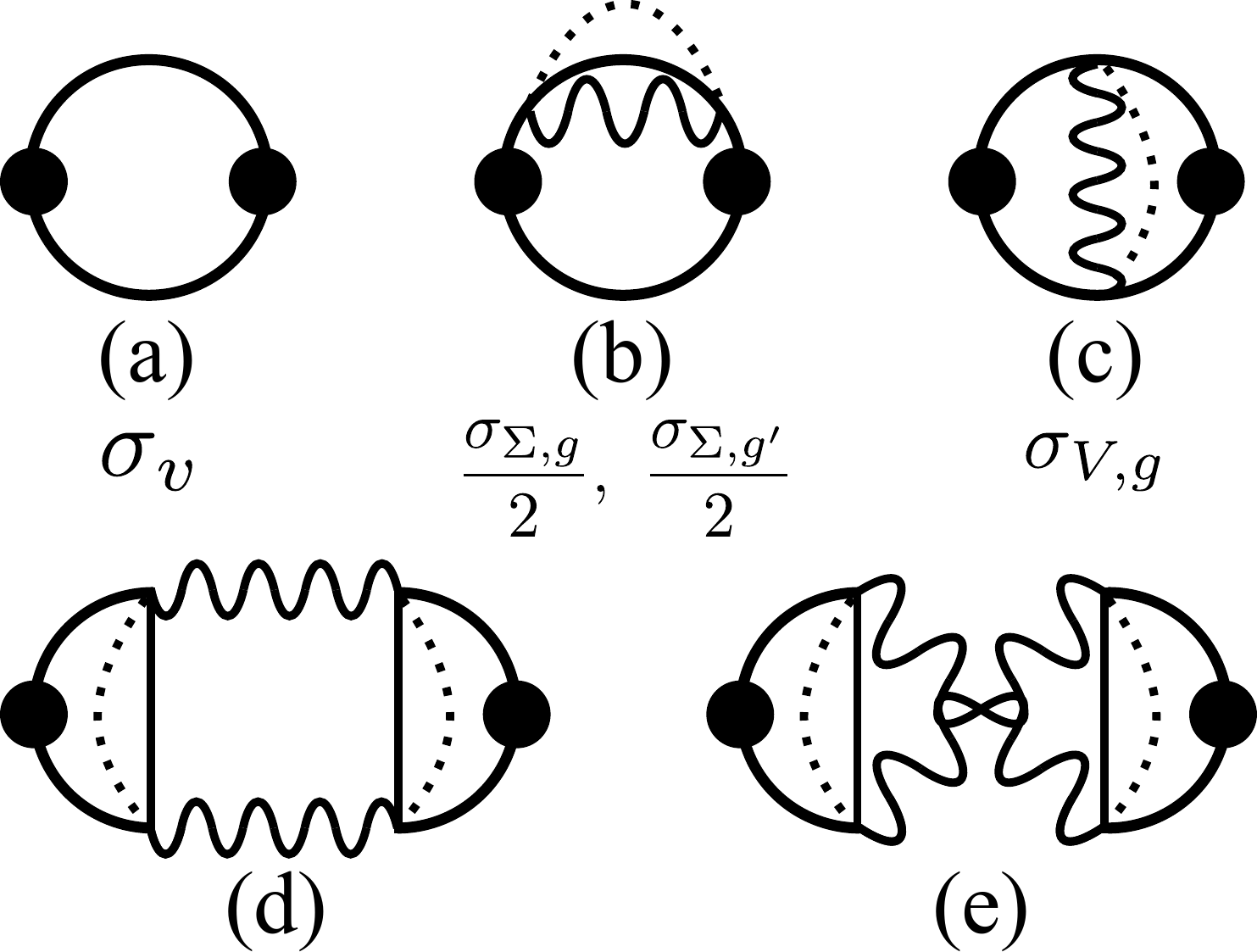}
\caption{Contribution of perturbative interaction corrections to the conductivity in the large $N$ limit. The current operators are denoted by solid circles, and the wavy lines denote boson propagators. Dashed lines denote random flavor averaging of the interaction couplings. The fermion Green's functions (solid lines) include the effects of the disordered potential ($v$), and the the quantum-critical boson propagators include the effects of damping due to interactions. Vertex corrections ((c)-(e)) contain only $g$ interactions, as the contributions from $g'$ interactions vanish due to the decoupling of the momentum integrals in the loops containing the external current operators. The sum of the two  ``Aslamazov-Larkin" diagrams ((d), (e)) vanishes exactly in the limit of large Fermi energy, and the perturbative result (\ref{eq:sigmavggp2}) is therefore valid to all orders in the interaction strength (Supplementary Information).
}
\label{fig:Kubo}
\end{figure}

{\it Potential disorder.} We now add a spatially random fermion potential 
\begin{align}
& \mathcal{S}_v =   \frac{1}{\sqrt{N}} \int d^2 r d \tau \, v_{ij} (\br)  \psi_i^{\dagger} (\br, \tau) \psi_j(\br, \tau) \nonumber \\
& \overline{v_{ij} (\br)} = 0 \,, \quad \overline{v^\ast_{ij} (\br) v_{lm} (\br')} = v^2 \, \delta(\br-\br') \delta_{il} \delta_{jm}
\end{align}
and here the overline is now an average over both spatial co-ordinates and flavor space. The large $N$ limit of the $G$-$\Sigma$ theory \cite{criticalII} yields a saddle point which has statistical translational invariance, and is
similar to earlier studies \cite{HLR,ChubukovMaslov,BergWang}. The low frequency boson propagator now has the diffusive form $\sim (q^2 + c_d |\omega|)^{-1}$ with $z=2$, while the fermion self energy has an elastic scattering term, along with a marginal Fermi liquid \cite{MFL89} inelastic term at low frequencies
\begin{align}
\Pi (i \omega, {\bm q}) & =-\frac{\mathcal{N} g^2|\omega|}{\Gamma}, \quad \Gamma = 2\pi v^2 \mathcal{N}, \\
\Sigma(i\omega, {\bm k} = k_F \hat{k}) & = - i \frac{\Gamma}{2} \mbox{sgn}(\omega) - \frac{ig^2 \omega}{2\pi^2 \Gamma}\ln\left(\frac{e \Gamma^3}{\mathcal{N} g^2 v_F^2 |\omega|}\right), \nonumber
\end{align}
at $T = 0$. However, the marginal Fermi liquid self energy, while leading to a $T\ln(1/T)$ specific heat, does {\it not\/} \cite{criticalII} lead to the claimed \cite{MFL89} linear-$T$ term in the DC resistivity, as it arises from forward scattering of electrons off the $\mathbf{q}\sim0$ bosons. These forward scattering processes are unable to relax either current or momentum due to the small wavevector of the bosons involved and the momentum conservation of the $g$ interactions. As a result, even a perturbative computation of the conductivity at $\mathcal{O}(g^2)$ (Fig.~\ref{fig:Kubo}) shows a cancellation  between the interaction-induced self energy contributions and the interaction-induced vertex correction, leading to a DC conductivity that is just a constant, set by the elastic potential disorder scattering rate $\Gamma$. The leading frequency dependence of the optical conductivity at frequencies $\omega\ll \Gamma$ is just a constant, and there is no linear in frequency correction \cite{criticalII}. Correspondingly, in the DC limit, there is no linear in $T$ correction, and a conventional $T^2$ correction is expected.

{\it Interaction disorder.} Our main results are obtained with additional spatially random interactions. In principle, such terms will be generated under a renormalization analysis from $\mathcal{S}_v$. However, such a renormalization is not part of our large $N$ limit, and so we account for such interactions by adding an explicit term:
\begin{align}
\label{eq:disint}
& \mathcal{S}_{{g'}} =  \frac{1}{N} \int d^2 r d\tau \, {g'}_{ijl} (\br) \psi_i^{\dagger} (\br, \tau) \psi_j(\br, \tau) \phi_l (\br, \tau)  \\ 
& \overline{{g'}_{ijl}(\br)} = 0 \, , \quad  \overline{{{g'}}^\ast_{ijl}(\br){g'}_{abc}(\br')} = {{g'}}^2\,\delta(\br-\br')\delta_{ia}\delta_{jb}\delta_{lc} \,. \nonumber
\end{align}
Note that $v$, $g$, and $g'$ are all independent flavor-random variables. Earlier work has considered the limiting case $g=0$, $v=0$, ${g'} \neq 0$ \cite{Altman1,Esterlis:2021eth}. We will instead describe the more physically relevant regime where spatial disorder is a weaker perturbation to a clean quantum-critical system, with $g$ the largest interaction coupling. We therefore now have all of $g,v,g'$ nonzero. The $G$-$\Sigma$ theory of $\mathcal{S}_g+\mathcal{S}_v+\mathcal{S}_{g'}$ is described in the supplement.

As with $\mathcal{S}_g+\mathcal{S}_v$ above, we find a statistical translational invariance at large $N$, with a low frequency boson propagator characterized by $z=2$, and the low frequency fermion self energy with an elastic scattering term, along with a marginal Fermi liquid inelastic term 
\footnote{Since $g'$ is a small fluctuation about $g$, we will consider $\mathcal{N}{g'}^2<g^2/\Gamma$, which makes the $g'$ contributions to $\Pi$ and $\Sigma$ smaller than the $g$ contributions in (\ref{eq:sigmavggp})} 
\begin{align}
\Pi (i \omega, {\bm q}) & =-\frac{\mathcal{N} g^2|\omega|}{\Gamma} - \frac{\pi}{2}\mathcal{N}^2{g'}^2|\omega| \equiv - c_d |\omega| ,\nonumber \\
\Sigma(i\omega, {\bm k} = k_F \hat{k}) & = - i \frac{\Gamma}{2} \mbox{sgn}(\omega) - \frac{ig^2 \omega}{2\pi^2 \Gamma}\ln\left(\frac{e \Gamma^2}{v_F^2 c_d |\omega|}\right)\nonumber \\
&-\frac{i\mathcal{N}{g'}^2\omega}{4\pi}\ln\left(\frac{e\Lambda_d^2}{c_d |\omega|}\right)~~(T=0),
\label{eq:sigmavggp}
\end{align}
where $\Lambda_d\sim\Gamma/v_F$. This self-energy leads to a $T \ln (1/T)$ specific heat, as for the large $g'$ case \cite{Esterlis:2021eth}. However, there is now an important difference with respect to the previous case where $g'=0$, which leads to markedly different charge transport properties: the marginal Fermi liquid self energy now contains a term (last line of \eqref{eq:sigmavggp}), that does {\it not} arise solely from forward scattering of electrons. This term is produced by the {\it disordered} part of the interactions in \eqref{eq:disint}. Therefore, this part of the self energy represents scattering that relaxes {\it both} current and momentum carried by the electron fluid, and therefore its imaginary part on the real frequency axis determines the inelastic {\it transport} scattering rate. 

We can show this as follows by computing the conductivity using the Kubo formula. If we work perturbatively in $g$ and $g'$, then the conductivity at $\mathcal{O}(g^2)$ and $\mathcal{O}(g'^2)$ in the large $N$ limit is given by the sum of self energy contributions and vertex corrections (Fig. \ref{fig:Kubo}). However, due to the isotropy of the scattering processes arising from the $g'$ interactions, only the vertex correction due to the $g$ interactions survives. The conductivity up to the first sub-leading frequency dependent correction is then given by 
(see Supplementary Information)
\begin{align}
&\frac{1}{N}\mathrm{Re}[\sigma(\omega \gg T)] =  \sigma_v + \sigma_{\Sigma,g} + \sigma_{V,g} + \sigma_{\Sigma,g'}; \nn
&\sigma_v(\omega) = \frac{\mathcal{N}v_F^2}{2\Gamma},~~\sigma_{\Sigma,g}(\omega) = - \frac{\mathcal{N}v_F^2g^2|\omega|}{8\pi\Gamma^3}, \nn
&\sigma_{V,g}(\omega) = \frac{\mathcal{N}v_F^2g^2|\omega|}{8\pi\Gamma^3},\sigma_{\Sigma,g'}(\omega) = - \frac{\mathcal{N}^2v_F^2{g'}^2|\omega|}{16\Gamma^2}.
\label{eq:sigmavggp1}
\end{align}
Note that the $g^2$ vertex and self-energy terms cancel, and we have
\begin{align}
N\mathrm{Re}\left[\frac{1}{\sigma(\omega \gg T)}\right] & = \frac{1}{\mathcal{N}v_F^2}\left[ 2\Gamma+\frac{\mathcal{N}{g'}^2|\omega|}{4} \right]\,.
\label{eq:sigmavggp2}
\end{align}
The ${g'}^2$ term does not cancel, and leads to a linear in frequency correction to the constant transport scattering rate $\Gamma$. In the opposite limit $|\omega| \ll T$, this translates into a $T$-linear correction to the resistivity in the DC limit; computing the co-efficient of the linear-$T$ resistivity requires a self-consistent 
numerical analysis, which has been carried out in the large $g'$ limit \cite{Altman1,Esterlis:2021eth}.
Remarkably, the slope of this scattering rate with respect to $|\omega|$ (and therefore $T$) does not depend on $\Gamma$ and hence on the residual ($\omega=T=0$) resistivity. In the Supplementary Information we show that the perturbative result described here continues to be valid under a full resummation of all diagrams at large $N$ in the Kubo formula, as all surviving higher order contributions are merely repetitions of the interaction insertions in Fig. \ref{fig:Kubo}b,c.

We can also consider the case where $v=0$ but $g\neq0$ and $g'\neq0$. In this case we have (at $T=0$) (Supplementary Information)
\begin{align}
\Pi(i\omega, \bq) &= -c_b \frac{|\omega|}{|\bq|} - \frac{\pi}{2}\mathcal{N}^2{g'}^2|\omega|, \\
\Sigma(i\omega,\mathbf{k}) &= -ic_f\mathrm{sgn}(\omega)|\omega|^{2/3} - \frac{i\mathcal{N}{g'}^2\omega}{6\pi}\ln\left(\frac{e\tilde{\Lambda}^3}{c_b|\omega|}\right), \nonumber
\label{eq:sigmaggp}
\end{align}
where $\tilde{\Lambda}\sim g^2/({g'}^2v_F\mathcal{N})$ is a UV momentum cutoff. Interestingly, the disordered interactions induce a marginal Fermi liquid term in $\Sigma$, which manifests as the first higher order correction to the translationally invariant result (\ref{eq:sigmag}) 
\footnote{Because the marginal Fermi liquid correction is sub-leading, the specific heat in this case is $\sim T^{2/3}$ \cite{Esterlis:2021eth} and not $\sim T\ln(1/T)$}.

It is sufficient in this $v=0$ but $g\neq0$ and $g'\neq0$ case to compute the conductivity using the theory of modes in the vicinity of antipodal points on the Fermi surface 
\footnote{Corrections to the conductivity that may arise from going beyond this antipodal patch theory are sub-leading to the effects of the disordered interactions}. 
We then find, as before, that $\sigma_{\Sigma,g}$ and $\sigma_{V,g}$ cancel, and (Supplementary Information)
\begin{align}
\frac{1}{N}\sigma(i\omega\gg T) &= \frac{\mathcal{N}v_F^2}{2\omega} - \frac{ \mathcal{N}^2v_F^2{g'}^2}{24\pi\omega}\ln\left(\frac{e^3\tilde{\Lambda}^6}{c_b^2\omega^2}\right) \nn
&\approx \frac{\mathcal{N}v_F^2}{2\omega+\displaystyle\frac{\mathcal{N}{g'}^2\omega}{6\pi}\ln\left(\frac{e^3\tilde{\Lambda}^6}{c_b^2\omega^2}\right)}; \\
\frac{1}{N}\frac{\mathrm{Re}[\sigma(\omega\gg T)]}{\mathcal{N}^2v_F^2{g'}^2} &= \Bigg[ 6|\omega|\Bigg[\left(2+\frac{\mathcal{N}{g'}^2}{6\pi}\ln\left(\frac{e^3\tilde{\Lambda}^6}{c_b^2\omega^2}\right)\right)^2 \nn
&~~~~~~~~+\frac{\mathcal{N}^2{g'}^4}{36}\Bigg]\Bigg]^{-1}. \nonumber
\end{align}
The transport scattering rate is therefore still linear in $|\omega|$ (and hence $T$), up to logarithms, and there is no residual resistivity when $v=0$ despite the presence of disorder in $g'$. This also turns out to be valid to all orders in perturbation theory in the large $N$ limit (Supplementary Information).

{\it Crossovers.} For energy ($E$) scales larger than $E_{c,1}\sim \Gamma^2/(v_F^2c_d)$, but smaller than $E_{c,2}\sim g^4/({g'}^6v_F^2\mathcal{N}^4)$ ($E_{c,1}<E_{c,2}$, because $\mathcal{N}{g'}^2<g^2/\Gamma$ as disorder is a correction to the clean system), the leading frequency dependence of the inelastic part of the fermion self energy induced by $g$ changes from $i\omega\ln(1/|\omega|)$ to $i\mathrm{sgn}(\omega)|\omega|^{2/3}$, as in the theory with $v=0$ described above (Supplementary Information). However then, as shown above for $v=0$, the $|\omega|$ or $T$ dependence of the transport scattering rate continues to arise from $g'$ and remains linear (up to logarithms), but with a slope that is a factor of $\sim2/3$ times the slope in the $E<E_{c,1}$ theory.

For energy scales larger than $E_{c,2}$, there is an additional crossover to the theory with $g=0$ considered in Refs. \cite{Altman1,Esterlis:2021eth}, which also has a linear $|\omega|$ or $T$ dependence (up to logarithms) of the transport scattering rate, but now with the same slope as in the $E<E_{c,1}$ theory (Supplementary Information).

{\it Planckian behavior.} Experimental analyses \cite{Gael21,Paschen22} have compared the slope of the linear-$T$ resistivity to the renormalization of the effective mass in a proximate Fermi liquid, and so deduced a `scattering time' $\tau_{\rm tr}^\ast$ appearing in a Drude formula for the resistivity. In our theory, we obtain
\beq
\frac{1}{\tau_{\rm tr}^\ast} = \alpha \, \frac{k_B T}{\hbar}\,.
\eeq
The dimensionless number $\alpha$ has been computed previously \cite{Altman1,Esterlis:2021eth} in the limit $g' \gg g$ to be $\alpha \approx (\pi/2) \times $(ratio of logarithms of $T$). For smaller $g'$ we find (at $v\neq0$) (Supplementary Information)
\begin{align}
\alpha &\approx \frac{\pi}{2}\frac{g'^2}{g'^2 L_1(T)+\displaystyle\frac{g^2}{\Gamma\mathcal{N}} L_2(T)},~~L_{1,2}(T)\sim -\ln T. 
\end{align}
Therefore, `Planckian behavior' ($\alpha\sim\mathcal{O}(1)$ and depending only slowly on $T$ and non-universal parameters) only occurs in the regime of large $g'$ considered in Refs. \cite{Altman1,Esterlis:2021eth}. Otherwise, $\alpha\ll 1$ when $g$ is the largest interaction coupling. Our theory therefore provides a concrete realization of the often conjectured ``Planckian bound" of $\alpha\lesssim 1$ on the transport scattering times of quantum-critical metals \cite{Hartnoll:2021ydi,Zaanen,Sankar2022}. It is worth noting that quantum-critical $T$-linear resistivity with $\alpha\ll 1$ has been recently observed in experiments on heavy fermion materials \cite{Paschen22}. Finally, for $v=0$ but $g\neq0$, $\alpha\ll 1$ and has a power-law dependence on $T$; therefore there is manifestly no Planckian behavior in this case.

{\it Scalar mass disorder.} Finally, we consider spatial disorder in the scalar `mass' $m_b$, and argue that it does not modify our results over substantial intermediate scales. Such a term is not allowed for emergent gauge fields, but it can appear as a fluctuation in the position of the quantum-critical point for the cases where $\phi$ is a symmetry breaking order parameter.
\beq
\mathcal{S}_{w} = \int d\tau \frac{1}{2\sqrt{N}} \int d^2 r \sum_{ij=1}^{N} w_{ij} (\br) \phi_{i}(\br,\tau) \phi_{j}(\br,\tau)
\eeq
with
\beq
\overline{w_{ij} (\br) w_{lm} (\br')} = \frac{w^2}{2} \, \delta(\br-\br') \left( \delta_{il} \delta_{jm} + \delta_{im} \delta_{jl} \right)
\eeq
The large $N$ analysis shows that $\mathcal{S}_{w}$ is strongly relevant, and so $w$ may well be the most significant source of spatial disorder in experimental systems. Consequently, it is appropriate to account for $\mathcal{S}_{w}$ first, by transforming to the bases of eigenmodes of $\phi$ which are eigenstates of the harmonic terms for $\phi$ in a given disorder realization. In this new basis, we will obtain a theory which has the same form as $\mathcal{S}_g + \mathcal{S}_{v} + \mathcal{S}_{{g'}}$ with additional spatial disorder in the couplings, including in $K$. However, it is not difficult to show that spatial disorder in $K$ is unimportant. So $\mathcal{S}_{w}$ can be absorbed in a renormalization of the values of $v$ and ${g'}$, and we can continue to use our results for $\mathcal{S}_g + \mathcal{S}_{v} + \mathcal{S}_{{g'}}$. A more thorough analysis of disorder fluctuation effects is required to determine if this transformation remains valid at the longest scales near the quantum critical point.

{\it Discussion.} ({\it i\/}) A phenomenologically attractive feature of our theory is that the residual resistivity and the slope of the linear-$T$ resistivity are determined by different types of disorder: respectively, the potential disorder $v$ (which determines the elastic scattering rate $\Gamma$) and the interaction disorder $g'$ (which determines the inelastic self energy in the last term of (\ref{eq:sigmavggp})).\\
({\it ii\/}) We obtain a marginal Fermi liquid electron self-energy \cite{MFL89}, as often observed in quantum-critical metals \cite{Paglione20}. \\
({\it iii\/}) As the coupling $g’$ is spatially random, momentum is not conserved at its Yukawa interaction vertex. The physical properties therefore remain unchanged
for order parameters at nonzero momentum, and also for theories with multiple Fermi surfaces.\\
({\it iv\/}) We could have computed all diagrams directly at $N=1$, and found the same crucial cancellations. The large $N$ mainly serves to systematically select a consistent set of diagrams to resum from the saddle point of an effective action. Furthermore, the large $N$/Eliashberg theory agrees well with quantum Monte Carlo (QMC) studies in the clean limit (carried out with the number of fermion/boson flavors of order one) \cite{Klein20,MengChubukov,BergFernandes,ChubukovKivelson}, and does not have a potentially destabilizing Schwarzian zero mode \cite{Esterlis:2021eth}. A  comparison with QMC for the disordered case requires significantly more advanced computational techniques, and is the subject of ongoing work \cite{Patel_BAPS}.\\
({\it v\/}) Our theory of the influence of spatial disorder includes some disorder terms to all orders, and this yields the $z=2$ diffusive scalar propagator. This is in contrast to the perturbative disorder analysis of earlier memory function treatments \cite{Hartnoll:2014gba,BergWang}.\\
({\it vi\/}) Unlike earlier approaches (see Ref. \cite{Chowdhury:2021qpy}) to constructing controlled theories of strongly correlated metals with low-temperature $T$-linear resistivity, there is no local criticality in our new theory. The quantum-critical scalar fluctuations live in two, and not zero spatial dimensions. \\
({\it vii\/}) When the values of the interaction couplings and $T$ are large enough to make the fermion self energy $\Sigma$ comparable to the Fermi energy, we expect the theories described here to cross over into a `bad metal' regime \cite{Patel2017}; it would be interesting to examine the transport properties of such a regime.

{\it Acknowledgements.} We thank Erez Berg, Andrey Chubukov, and Joerg Schmalian for valuable discussions. H.G. and S.S. were supported by the National Science Foundation under Grant No.~DMR-2002850.  A.A.P. was supported by the Miller Institute for Basic Research in Science. This work was also supported by the Simons Collaboration on Ultra-Quantum Matter, which is a grant from the Simons Foundation (651440, S.S.). H.G. was supported in part by the Heising-Simons Foundation, the Simons Foundation, and National Science Foundation Grant No. NSF PHY-1748958. The Flatiron Institute is a division of the Simons Foundation.

\bibliography{refs.bib}

\ifarXiv
    

\section*{Supplementary material}

# Supplementary Information for
# Universal theory of strange metals from spatially random interactions

Aavishkar A. Patel

*Department of Physics, University of California Berkeley, Berkeley CA 94720, USA and*
*Center for Computational Quantum Physics, Flatiron Institute, New York, New York, 10010, USA*

Haoyu Guo

*Department of Physics, Harvard University, Cambridge MA 02138, USA and*
*Kavli Institute for Theoretical Physics, University of California, Santa Barbara, California 93106, USA*

Ilya Esterlis

*Department of Physics, Harvard University, Cambridge MA 02138, USA*

Subir Sachdev

*Department of Physics, Harvard University, Cambridge MA 02138, USA and*
*School of Natural Sciences, Institute for Advanced Study, Princeton, NJ-08540, USA*

**CONTENTS**

This supplement will consider the most general model, including the translationally invariant Yukawa coupling $g$, the potential disorder $v$, and an additional spatial randomness in the Yukawa coupling $g'$. The Green's functions of the model with only $g \neq 0$ were described in Ref. [1], and the transport properties of this model were discussed in Ref. [2]: this model has zero resistivity in the absence of umklapp, because of momentum conservation. The case with $v \neq 0$ and $g \neq 0$ was described in Ref. [2]: this model has a marginal Fermi liquid self energy for the fermions, but a Fermi liquid-like resistivity. Here we will describe a (mostly) self-contained description of the most general case with $v \neq 0$, $g \neq 0$, and $g' \neq 0$: we will show that this model has a marginal Fermi liquid self energy for the fermions, along with a strange metal linear-$T$ resistivity.

## SI. ACTION

We recall the action for all terms, and describe the associated large $N$ saddle point. We start with the Lagrangian for the critical Fermi surface without disorder

$$\mathcal{L} = \sum_i \psi_i^\dagger(\partial_\tau + \varepsilon_{\mathbf{k}} - \mu)\psi_i + \frac{1}{2}\sum_i \phi_i(-\partial_\tau^2 + \omega_{\mathbf{q}}^2 + m_b^2)\phi_i + \sum_{ijl} \frac{g_{ijl}}{N}\psi_i^\dagger \psi_j \phi_l, \tag{S1}$$

where $g_{ijl}$ is spatially independent,

$$\overline{|g_{ijl}|^2} = g^2. \tag{S2}$$

This was the model studied in Refs. [1, 2].

To this we add a random potential coupling to the fermions

$$S_v = \int d\tau \frac{1}{\sqrt{N}} \sum_{\mathbf{r}} \sum_{ij=1}^N v_{ij}(\mathbf{r})\psi_i^\dagger(\mathbf{r},\tau)\psi_j(\mathbf{r},\tau), \tag{S3}$$

where $r$ labels lattice sites. The potential $v_{ij}(\mathbf{r})$ is random both in position and flavor space

$$\overline{v_{ij}^*(\mathbf{r})v_{lm}(\mathbf{r}')} = v^2\,\delta(\mathbf{r}-\mathbf{r}')\delta_{il}\delta_{jm}, \tag{S4}$$

This was studied in Ref. [2]. We can also add a similar random potential coupling to the bosons

$$S_w = \int d\tau \frac{1}{2\sqrt{N}} \sum_{\mathbf{r}} \sum_{ij=1}^N w_{ij}(\mathbf{r})\phi_i(\mathbf{r},\tau)\phi_j(\mathbf{r},\tau), \tag{S5}$$

with

$$\overline{w_{ij}(\mathbf{r})w_{lm}(\mathbf{r}')} = \frac{w^2}{2}\,\delta(\mathbf{r}-\mathbf{r}')\left(\delta_{il}\delta_{jm} + \delta_{im}\delta_{jl}\right). \tag{S6}$$

The large $N$ saddle point equations for $S + S_v + S_w$ are now

$$\begin{aligned}
\Sigma(\tau,\mathbf{r}) &= g^2 D(\tau,\mathbf{r})G(\tau,\mathbf{r}) + v^2 G(\tau,\mathbf{r}=0)\delta^2(\mathbf{r}), \\
\Pi(\tau,\mathbf{r}) &= -G(-\tau,-\mathbf{r})G(\tau,\mathbf{r}) + w^2 D(\tau,\mathbf{r}=0)\delta^2(\mathbf{r}), \\
G(i\omega,\mathbf{k}) &= \frac{1}{i\omega - \varepsilon_{\mathbf{k}} + \mu - \Sigma(i\omega,\mathbf{k})}, \\
D(i\Omega,\mathbf{q}) &= \frac{1}{\Omega^2 + \omega_{\mathbf{q}}^2 + m_b^2 - \Pi(i\Omega,\mathbf{q})}.
\end{aligned} \tag{S7}$$

We choose the conventional dispersions $\varepsilon_{\mathbf{k}} = |\mathbf{k}|^2/(2m)$ and $\omega_{\mathbf{q}}^2 = |\mathbf{q}|^2 = q^2$, and tune the system to the quantum-critical point with $m_b^2 - \Pi(0,0) = 0$ at $T=0$. The $v^2$ term leads to an impurity scattering lifetime $\sim iv^2\mathrm{sgn}(\omega)$ in the fermion self energy. The $w^2$ term leads to an impurity scattering term $\sim w^2\ln(|\omega|)$ in the boson self energy. While this term is strongly relevant, we argue in the main text that we can neglect it following a rotation of the boson basis, after which it only ends up generating the $g'$ term.

Finally, we add the crucial disorder to the interaction term:

$$S_{g'} = \int d\tau \frac{1}{N} \sum_{\mathbf{r}} \sum_{ilj=1}^N g'_{ijl}(\mathbf{r})\psi_i^\dagger(\mathbf{r},\tau)\psi_j(\mathbf{r},\tau)\phi_l(\mathbf{r},\tau), \tag{S8}$$

where $g'_{ijl}(\mathbf{r})$ is spatially random,

$$\overline{g'^*_{ijl}(\mathbf{r})g'_{abc}(\mathbf{r}')} = g'^2\delta(\mathbf{r}-\mathbf{r}')\delta_{ia}\delta_{jb}\delta_{lc}. \tag{S9}$$

For $S_{\text{all}} = S + S_v + S_w + S_{g'}$, after averaging over disorder, we obtain the $G$-$\Sigma$-$D$-$\Pi$ action which generalizes those in Refs. [1, 2]:

$$\begin{aligned}
\frac{S_{\text{all}}}{N} &= -\ln\det(\partial_\tau + \varepsilon_{\mathbf{k}} - \mu + \Sigma) + \frac{1}{2}\ln\det(-\partial_\tau^2 + \omega_{\mathbf{q}}^2 + m_b^2 - \Pi) \\
&+ \int d\tau d^2r \int d\tau' d^2r' \left[ -\Sigma(\tau',\mathbf{r}';\tau,\mathbf{r})G(\tau,\mathbf{r};\tau',\mathbf{r}') + \frac{1}{2}\Pi(\tau',\mathbf{r}';\tau,\mathbf{r})D(\tau,\mathbf{r};\tau',\mathbf{r}') \right. \\
&+ \frac{g^2}{2}G(\tau,\mathbf{r};\tau',\mathbf{r}')G(\tau',\mathbf{r}';\tau,\mathbf{r})D(\tau,\mathbf{r};\tau',\mathbf{r}') + \frac{v^2}{2}G(\tau,\mathbf{r};\tau',\mathbf{r}')G(\tau',\mathbf{r}';\tau,\mathbf{r})\delta(\mathbf{r}-\mathbf{r}') \\
&\left. - \frac{w^2}{2}D(\tau,\mathbf{r};\tau',\mathbf{r}')D(\tau',\mathbf{r}';\tau,\mathbf{r})\delta(\mathbf{r}-\mathbf{r}') + \frac{g'^2}{2}G(\tau,\mathbf{r};\tau',\mathbf{r}')G(\tau',\mathbf{r}';\tau,\mathbf{r})D(\tau,\mathbf{r};\tau',\mathbf{r}')\delta(\mathbf{r}-\mathbf{r}') \right]
\end{aligned} \tag{S10}$$

The path integral over $G$-$\Sigma$-$D$-$\Pi$ with the action in (S10) is our *universal theory of strange metals*. Note that the Green's functions and self energies in (S10) are bi-local in space and time, and the terms involving an average over spatial disorder have a $\delta(\mathbf{r}-\mathbf{r}')$ in them because of the local nature of the disorder. At the large $N$ saddle point, the Green's functions and self energies become functions only of time and spatial differences, because the disorder average restores a statistical spatial translational symmetry, In this manner, we obtain the most general saddle point equations:

$$\begin{aligned}
\Sigma(\tau,\mathbf{r}) &= g^2 D(\tau,\mathbf{r})G(\tau,\mathbf{r}) + v^2 G(\tau,\mathbf{r}=0)\delta^2(\mathbf{r}) + g'^2 G(\tau,\mathbf{r}=0)D(\tau,r=0)\delta^2(\mathbf{r}), \\
\Pi(\tau,\mathbf{r}) &= -g^2 G(-\tau,-\mathbf{r})G(\tau,\mathbf{r}) + w^2 D(\tau,\mathbf{r}=0)\delta^2(\mathbf{r}) - g'^2 G(-\tau,\mathbf{r}=0)G(\tau,\mathbf{r}=0)\delta^2(\mathbf{r}), \\
G(i\omega,\mathbf{k}) &= \frac{1}{i\omega - \varepsilon_{\mathbf{k}} + \mu - \Sigma(i\omega,\mathbf{k})}, \\
D(i\Omega,\mathbf{q}) &= \frac{1}{\Omega^2 + \omega_{\mathbf{q}}^2 + m_b^2 - \Pi(i\Omega,\mathbf{q})}.
\end{aligned} \tag{S11}$$

## SII. SELF ENERGIES

We take $w=0$ and focus on the critical point. We then compute the self-consistent solutions for the fermion and boson Green's functions. We first consider the case of nonzero $v$, at low frequencies and $T=0$. For the internal lines in the self-energy diagrams, because the interaction contribution to the fermion self energy and the bare $i\omega$ term are both much smaller than the impurity scattering rate at low frequencies, we can approximate $G(i\omega,\mathbf{k}) \simeq G_0(i\omega,\mathbf{k}) = 1/(i\Gamma\text{sgn}(\omega)/2 - v_F k)$, where $v_F$ is the Fermi velocity, $k = |\mathbf{k}| - k_F = |\mathbf{k}| - \sqrt{2m\mu}$, and $\Gamma = v^2 k_F/v_F = 2\pi v^2 \mathcal{N}$ is the impurity scattering rate. Then, we obtain, for the boson self energy at criticality:

$$\Pi(i\Omega,\mathbf{q}) - \Pi(0,0) = \Pi_g(i\Omega,\mathbf{q}) + \Pi_{g'}(i\Omega)$$

$$\Pi_g(i\Omega,\mathbf{q}) = -g^2 k_F \int_{-\infty}^{\infty} \frac{d\omega}{2\pi} \int_{-\pi}^{\pi} \frac{d\theta}{2\pi} \int_{-\infty}^{\infty} \frac{dk}{2\pi} \frac{1}{i\frac{\Gamma}{2}\text{sgn}(\omega) - v_F k} \frac{1}{i\frac{\Gamma}{2}\text{sgn}(\omega+\Omega) - v_F(k+q\cos\theta)}$$

$$= -\frac{g^2 k_F |\Omega|}{2\pi v_F \sqrt{\Gamma^2 + v_F^2 q^2}} \simeq -\frac{g^2 k_F |\Omega|}{2\pi v_F \Gamma} \equiv -\frac{\mathcal{N}g^2}{\Gamma}|\Omega|, \quad (v_F q \ll \Gamma)$$

$$\Pi_{g'}(i\Omega) = -g'^2 k_F^2 \int_{-\infty}^{\infty} \frac{d\omega}{2\pi} \int_{-\infty}^{\infty} \frac{dk}{2\pi} \int_{-\infty}^{\infty} \frac{dk'}{2\pi} \frac{1}{i\frac{\Gamma}{2}\text{sgn}(\omega) - v_F k} \left( \frac{1}{i\frac{\Gamma}{2}\text{sgn}(\omega+\Omega) - v_F k'} - \frac{1}{i\frac{\Gamma}{2}\text{sgn}(\omega) - v_F k'} \right)$$

$$= -\frac{g'^2 k_F^2 |\Omega|}{8\pi v_F^2} \equiv -\frac{\pi}{2}\mathcal{N}^2 g'^2 |\Omega|,$$

$$\Pi(i\Omega, \mathbf{q}) - \Pi(0,0) = -c_d |\Omega|. \tag{S12}$$

Here, we have assumed that the Fermi surface is circular, $k_F$ is large so that $k, q \ll k_F$ and the Fermi energy is essentially infinite. The limit of large Fermi energy is therefore translated into a limit of large $k_F$, while keeping $\mathcal{N}$ fixed. This is completely equivalent to taking the limit of large Fermi energy in other ways, such as by fixing $k_F$ and making $\mathcal{N}$ small, which would be appropriate for a lattice. In such cases, one would have to adjust the values of $v$ and $g'$ so that $\Gamma = 2\pi \mathcal{N} v^2$ and $\mathcal{N} g'^2$ do not become arbitrarily large or arbitrarily small, which is not a problem.

At $T \neq 0$, we would expect a boson thermal mass $m_b^2(T) = m_b^2 - \Pi(0,0) \sim T/\ln(1/T)$ due to the marginal scaling dimension of the boson self interaction, since the dynamical critical exponent of the boson is $z = 2$ and its spatial dimensionality is $d = 2$ [1, 3].

We now consider the fermion self energy due to interactions $\Sigma(i\omega, \mathbf{k})$, which we also split into clean and momentum-independent disordered contributions $\Sigma(i\omega, \mathbf{k}) = \Sigma_g(i\omega, \mathbf{k}) + \Sigma_{g'}(i\omega)$. The disordered contribution $\Sigma_{g'}(i\omega)$ is straightforward to compute:

$$\Sigma_{g'}(i\omega) = g'^2 k_F \int_{-\infty}^{\infty} \frac{dk}{2\pi} \int_0^{\infty} \frac{qdq}{2\pi} \int_{-\infty}^{\infty} \frac{d\Omega}{2\pi} \frac{1}{i\frac{\Gamma}{2}\text{sgn}(\omega+\Omega) - v_F k} \frac{1}{q^2 + c_d |\Omega|}$$

$$\simeq -\frac{ig'^2 k_F \omega}{8\pi^2 v_F} \ln\left(\frac{e\Lambda_d^2}{c_d |\omega|}\right) \equiv -\frac{i\mathcal{N} g'^2 \omega}{4\pi} \ln\left(\frac{e\Lambda_d^2}{c_d |\omega|}\right), \tag{S13}$$

where $\Lambda_d \sim \Gamma/v_F$ is a UV cutoff on $q$. This part of the self energy corresponds to current and momentum relaxing scattering induced by the spatially random interactions, and has a marginal Fermi liquid form. These expressions also imply a frequency cutoff of $\Gamma^2/(v_F^2 c_d)$ on the low energy theory.

The computation of $\Sigma_g(i\omega, \mathbf{k})$ is a little more involved, and unlike $\Sigma_{g'}$, it has some momentum dependence. It is given by

$$\Sigma_g(i\omega, \mathbf{k}) = g^2 \int \frac{d^2\mathbf{k}'}{(2\pi)^2} \int_{-\infty}^{\infty} \frac{d\omega'}{2\pi} G_0(i\omega', \mathbf{k}') D(i\omega - i\omega', \mathbf{k} - \mathbf{k}'). \tag{S14}$$

For the computation of the scaling of the real part of the conductivity, the momentum dependence of $\Sigma_c(i\omega, \mathbf{k})$ does not matter, and one can use its value for $\mathbf{k}$ exactly on the Fermi surface, which is given by

$$\Sigma_g(i\omega, \mathbf{k} = k_F \hat{k}) \simeq -\frac{ig^2 \omega}{2\pi^2 \Gamma} \ln\left(\frac{e\Gamma^2}{v_F^2 c_d |\omega|}\right). \tag{S15}$$

Note that this also has a marginal Fermi liquid form, but doesn't correspond to current and momentum relaxing scattering, and therefore will not contribute to transport, as we will show later. For completeness, we provide the derivation of momentum dependent expression as well. Eq. (S14) may be expressed as

$$\Sigma_g(i\omega, \mathbf{k}) = g^2 k_F \int_{-\infty}^{\infty} \frac{dk'}{2\pi} \int_{-\pi}^{\pi} \frac{d\theta}{2\pi} \int_{-\infty}^{\infty} \frac{d\omega'}{2\pi} \frac{1}{i\frac{\Gamma}{2}\text{sgn}(\omega') - v_F k'} \Big((k_F + k)^2 + (k_F + k')^2$$

$$- 2(k_F + k)(k_F + k')\cos\theta + c_d |\omega - \omega'|\Big)^{-1}$$

$$\simeq g^2 k_F \int_{-\infty}^{\infty} \frac{dk'}{2\pi} \int_{-\pi}^{\pi} \frac{d\theta}{2\pi} \int_{-\infty}^{\infty} \frac{d\omega'}{2\pi} \frac{1}{i\frac{\Gamma}{2}\text{sgn}(\omega') - v_F k'} \left(2k_F^2 + k^2 + k'^2 - 2(k_F^2 + kk')\cos\theta\right)^{-1}$$

$$\simeq \frac{g^2}{2} \int_{-\infty}^{\infty} \frac{dk'}{2\pi} \int_{-\infty}^{\infty} \frac{d\omega'}{2\pi} \frac{1}{i\frac{\Gamma}{2}\text{sgn}(\omega') - v_F k'} \frac{1}{\sqrt{(k-k')^2 + c_d |\omega - \omega'|}}. \tag{S16}$$

In the above, we neglected non-singular (as $k, \omega \to k', \omega'$) terms in the angle-integrated boson propagator, which are also additionally suppressed by additional powers of $1/k_F$. These will only contribute to $\Sigma_g(i\omega, \mathbf{k})$ at higher orders in $\omega$ beyond the marginal Fermi liquid form and are therefore not of interest to us. We can split $1/(i\Gamma \text{sgn}(\omega')/2 - v_F k')$ into real and imaginary parts. Then,

$$-\frac{v_F g^2}{2} \int_{-\infty}^{\infty} \frac{dk'}{2\pi} \int_{-\infty}^{\infty} \frac{d\omega'}{2\pi} \frac{k'}{\frac{\Gamma^4}{4} + v_F^2 k'^2} \frac{1}{\sqrt{(k-k')^2 + c_d|\omega - \omega'|}}$$

$$= -\frac{v_F g^2}{2} \int_{-\infty}^{\infty} \frac{dk'}{2\pi} \int_{-\infty}^{\infty} \frac{d\omega'}{2\pi} \frac{k'}{\frac{\Gamma^2}{4} + v_F^2 k'^2} \left( \frac{1}{\sqrt{(k-k')^2 + c_d|\omega - \omega'|}} - \frac{1}{\sqrt{c_d|\omega - \omega'|}} \right)$$

$$= -\frac{v_F g^2}{\pi c_d} \int_{-\infty}^{\infty} \frac{dk'}{2\pi} \frac{k'|k-k'|}{\frac{\Gamma^2}{4} + v_F^2 k'^2}$$

$$\simeq \frac{g^2 k \ln\left(1 + \frac{4v_F^2 \Lambda_d^2}{\Gamma^2}\right)}{2\pi^2 v_F c_d}, \quad \Lambda_d \sim \frac{\Gamma}{v_F}. \tag{S17}$$

This term is frequency independent, and therefore cannot lead to any dissipation relevant for transport; we therefore drop it as it is only a small ($\mathcal{O}(1/k_F)$) renormalization of the Fermi velocity. Then we have

$$\Sigma_g(i\omega, \mathbf{k}) \simeq -\frac{ig^2}{4} \int_{-\infty}^{\infty} \frac{dk'}{2\pi} \int_{-\infty}^{\infty} \frac{d\omega'}{2\pi} \frac{\Gamma \text{sgn}(\omega')}{\frac{\Gamma^2}{4} + v_F^2 k'^2} \frac{1}{\sqrt{(k-k')^2 + c_d|\omega - \omega'|}}$$

$$= -\frac{ig^2}{2\pi c_d} \int_{-\infty}^{\infty} \frac{dk'}{2\pi} \frac{\Gamma \text{sgn}(\omega)}{\frac{\Gamma^2}{4} + v_F^2 k'^2} \left( \sqrt{(k-k')^2 + c_d|\omega|} - |k-k'| \right)$$

$$= \frac{g^2 \text{sgn}(\omega)}{8\pi^2 c_d v_F^2} \left( 2i\Gamma \ln\left(\frac{c_d v_F^2 |\omega|}{\Gamma^2 + 4k^2 v_F^2}\right) \right.$$

$$+ \sqrt{c_d v_F^2 |\omega| + \left(kv_F + i\frac{\Gamma}{2}\right)^2} \left( \ln\left(\sqrt{c_d v_F^2 |\omega| + \left(kv_F + i\frac{\Gamma}{2}\right)^2} + i\frac{\Gamma}{2} + kv_F\right) \right.$$

$$- \ln\left(\sqrt{c_d v_F^2 |\omega| + \left(kv_F + i\frac{\Gamma}{2}\right)^2} - i\Gamma - kv_F\right) + 2\tanh^{-1}\left(\frac{kv_F + i\frac{\Gamma}{2}}{\sqrt{c_d v_F^2 |\omega| + \left(kv_F + i\frac{\Gamma}{2}\right)^2}}\right) \right)$$

$$+ \sqrt{c_d v_F^2 |\omega| + \left(kv_F - i\frac{\Gamma}{2}\right)^2} \left( \ln\left(\sqrt{c_d v_F^2 |\omega| + \left(kv_F - i\frac{\Gamma}{2}\right)^2} + i\frac{\Gamma}{2} - kv_F\right) \right.$$

$$- \ln\left(\sqrt{c_d v_F^2 |\omega| + \left(kv_F - i\frac{\Gamma}{2}\right)^2} - i\frac{\Gamma}{2} + kv_F\right) + 2\tanh^{-1}\left(\frac{-kv_F + i\frac{\Gamma}{2}}{\sqrt{c_d v_F^2 |\omega| + \left(kv_F - i\frac{\Gamma}{2}\right)^2}}\right) \right)$$

$$\left. + 2kv_F \left( \ln\left(-kv_F + i\frac{\Gamma}{2}\right) - \ln\left(kv_F + i\frac{\Gamma}{2}\right) + 2i\tan^{-1}\left(\frac{2kv_F}{\Gamma}\right) \right) \right). \tag{S18}$$

The momentum dependent corrections to (S15) induced by expanding (S18) in $k$ will not produce any dissipative contributions to the conductivity as we shall show in the next section, and therefore may be ignored. Furthermore, because $|\Sigma(i\omega, \mathbf{k})| \ll \Gamma$ at small frequencies, the above fermion and boson self-energies lead to a self-consistent solution of (S11) low frequencies.

When $v = 0$, $\Gamma = 0$, and therefore $\Pi_g(i\Omega, \mathbf{q}) = -c_b|\omega|/|\mathbf{q}|$ from (S12) [1]. $\Pi_{g'}(i\Omega)$, on the other hand, stays the same as in (S12), as the value of the fermion Green's function integrated over momentum does not depend on $\Gamma$, due to the fermion bandwidth being large. Since $\Pi_{g'} \ll \Pi_g$ then at small $(\omega, \mathbf{q})$, we can neglect $\Pi_{g'}$ when computing

the leading contributions to $\Sigma_g$ and $\Sigma_{g'}$. We then have $\Sigma_g(i\omega, \mathbf{k}) \simeq -ic_f \mathrm{sgn}(\omega)|\omega|^{2/3}$ [1], and

$$\begin{aligned}\Sigma_{g'}(i\omega) &\simeq {g'}^2 k_F \int_{-\infty}^{\infty} \frac{dk}{2\pi} \int_0^{\infty} \frac{qdq}{2\pi} \int_{-\infty}^{\infty} \frac{d\Omega}{2\pi} \frac{G(i(\omega+\Omega), k)}{q^2 + c_b \frac{|\Omega|}{q}} \\ &\simeq -i\pi\mathcal{N}{g'}^2 \int_0^{\infty} \frac{qdq}{2\pi} \int_{-\infty}^{\infty} \frac{d\Omega}{2\pi} \frac{\mathrm{sgn}(\omega+\Omega)}{q^2 + c_b \frac{|\Omega|}{q}} \\ &\simeq -\frac{i\mathcal{N}{g'}^2 \omega}{6\pi} \ln\left(\frac{e\tilde{\Lambda}^3}{c_b|\omega|}\right),\end{aligned} \tag{S19}$$

where $\tilde{\Lambda}$ is a UV cutoff on $q$. This part of the self energy corresponds to current and momentum relaxing scattering induced by the spatially random interactions, and has a marginal Fermi liquid form. The corrections coming from $\Pi_{g'}$ that we neglected here can be shown to be a non-dissipative $\mathcal{O}(i\omega)$ term to $\Sigma_g$ and a $\mathcal{O}(i\mathrm{sgn}(\omega)|\omega|^{4/3})$ term to $\Sigma_{g'}$ respectively, both of which can be neglected when compared to the above results for $\Sigma_g$ and $\Sigma_{g'}$. These results are also self-consistent at low energies, since $|\Sigma_{g'}| \ll |\Sigma_g|$, and $\Sigma_g$ is essentially the same as its self-consistent value in the theory with $g' = 0$ [1].

As discussed earlier in this section, the low energy solutions for $v \neq 0$ derived above are valid below a frequency scale $E_{c,1} \sim \Gamma^2/(v_F^2 c_d)$ and a corresponding momentum scale $\Lambda_d \sim \Gamma/v_F$. Above these scales, the $v_F^2 q^2$ term in the denominator of the expression for $\Pi_g(i\Omega, \mathbf{q})$ (S12) is no longer negligible compared to the $\Gamma^2$ term, and there is a crossover to $z = 3$ boson dynamics, with $\Pi_g(i\Omega, \mathbf{q}) = c_b|\omega|/|\mathbf{q}|$. $\Pi_{g'}(i\Omega)$, on the other hand, stays the same as given by (S12), as the value of the fermion Green's function integrated over momentum does not change under this crossover, due to the fermion bandwidth being large. Similar to the case of $v = 0$, we then have $\Sigma_g(i\omega, \mathbf{k}) \simeq -ic_f \mathrm{sgn}(\omega)|\omega|^{2/3}$, and $\Sigma_{g'}$ is given by (S19), with the fermion Green's function $G(i\omega, \mathbf{k}) \simeq 1/(i\Gamma \mathrm{sgn}(\omega)/2 - v_F k - \Sigma_g(i\omega, \mathbf{k}) - \Sigma_{g'}(i\omega))$. For frequency scales larger than $E_{c,2} \sim g^4/({g'}^6 v_F^2 \mathcal{N}^4) > E_{c,1}$, we can see that $\Sigma_{g'}$ dominates over $\Sigma_g$ (and $\Pi_{g'}$ also dominates over $\Pi_g$, after applying the momentum scaling $q^3 \sim c_b|\omega|$), which produces another crossover to an effective theory having $v \neq 0$, $g = 0$, and $g' \neq 0$. Such a theory has $\Sigma = \Sigma_{g'}$ given by (S13) [1, 3]. The momentum scale $\tilde{\Lambda}$ (which serves as the momentum cutoff in (S19)) corresponding to $E_{c,2}$ is therefore $\tilde{\Lambda} \sim c_b^{1/3} E_{c,2}^{1/3} \sim g^2/({g'}^2 v_F \mathcal{N})$.

## SIII. TRANSPORT PROPERTIES

We now proceed to compute the conductivity at $T = 0$ as a function of frequency $\omega$ for the models with disordered interactions. We first focus on the case with $v, g, g'$ all nonzero. Because the thermal mass of the $z = 2$ bosons (that is induced by boson self-interactions) scales as $T$ up to logarithms [1, 3, 4], this leads to $\omega/T$ scaling in $\mathrm{Im}[\Sigma_R(\omega, \mathbf{k})]$ up to logarithms [1, 3], in turn leading to $T$-linear resistivity. The last fact was explicitly demonstrated earlier in models with $g = 0$ [1, 3], and we will show that it continues to be valid here when $g \neq 0$.

To demonstrate the existence of const.$+T$-linear resistivity, or a const.$+\omega$-linear scattering rate at $T = 0$ in the real part of the conductivity, it is sufficient to look at the computation of the conductivity perturbatively in the interactions $g$ and $g'$, while keeping diagrams that are order $N$. The full ladder computation performed in Ref. [2] incorporates essentially the same principles and cancellations; for the sake of simplicity and to highlight the essential features of the calculation, we will describe the generalization to the ladder summation after the perturbative computation.

The non-interacting contribution to the real part of the $T = 0$ conductivity is trivially given by computing the

current-current correlation function (Fig. 1a of the main text)

$$\frac{1}{N}\text{Re}[\sigma_v(\omega)] = \frac{\text{Im}[(f_v(i\omega) - f_v(0))_{i\omega \to \omega + i0^+}]}{\omega},$$

$$f_v(i\omega) = -v_F^2 k_F \int_{-\infty}^{\infty} \frac{d\omega'}{2\pi} \int_{-\pi}^{\pi} \frac{d\theta \cos^2\theta}{2\pi} \int_{-\infty}^{\infty} \frac{dk}{2\pi} \frac{1}{i\frac{\Gamma}{2}\text{sgn}(\omega + \omega') - v_F k} \frac{1}{i\frac{\Gamma}{2}\text{sgn}(\omega) - v_F k},$$

$$\frac{1}{N}\text{Re}[\sigma_v(\omega)] = \frac{\mathcal{N} v_F^2}{2\Gamma}. \quad (S20)$$

Here, we have noted that $k_F$ is large, and therefore approximated $v_{\mathbf{k}} = v_F \hat{\mathbf{k}}$.

The perturbative contribution to the conductivity from the momentum-independent $\Sigma_{g'}(i\omega)$ in Fig. 1b of the main text is also straightforwardly computed at small frequencies:

$$\frac{1}{N}\text{Re}[\sigma_{\Sigma,g'}(\omega)] = \frac{\text{Im}[(f_{\Sigma,g'}(i\omega) - f_{\Sigma,g'}(0))_{i\omega \to \omega + i0^+}]}{\omega}$$

$$f_{\Sigma,g'}(i\omega) = -2v_F^2 k_F \int_{-\infty}^{\infty} \frac{d\omega'}{2\pi} \int_{-\pi}^{\pi} \frac{d\theta \cos^2\theta}{2\pi} \int_{-\infty}^{\infty} \frac{dk}{2\pi} \frac{1}{i\frac{\Gamma}{2}\text{sgn}(\omega + \omega') - v_F k} \frac{1}{\left(i\frac{\Gamma}{2}\text{sgn}(\omega') - v_F k\right)^2} \Sigma_{g'}(i\omega)$$

$$= \frac{\mathcal{N}^2 v_F^2 g'^2 \omega^2 \ln\left(\frac{e^3 \Lambda^4}{c_d^2 \omega^2}\right)}{16\pi \Gamma^2},$$

$$\frac{1}{N}\text{Re}[\sigma_{\Sigma,g'}(\omega)] = -\frac{\mathcal{N}^2 v_F^2 g'^2 |\omega|}{16 \Gamma^2}. \quad (S21)$$

The perturbative correction to the real part of the conductivity in Fig. 1b of the main text from $\Sigma_g(i\omega, \mathbf{k})$ may be computed using (S15). The computation then parallels (S21), and we obtain

$$\frac{1}{N}\text{Re}[\sigma_{\Sigma,g}(\omega)] = -\frac{\mathcal{N} v_F^2 g^2 |\omega|}{8\pi \Gamma^3}. \quad (S22)$$

If we use the momentum dependent form of $\Sigma_g(i\omega, \mathbf{k})$ instead, i.e., (S18) and then numerically perform the $k, \omega'$ integrals in the Kubo formula, we only obtain an analytic $\propto \omega^2$ correction to the current correlation function $f_{\Sigma,g}(i\omega) - f_{\Sigma,g}(0)$, over the result utilizing (S15) for the same. Therefore, $\text{Im}[(f_{\Sigma,g}(i\omega) - f_{\Sigma,g}(0))_{i\omega \to \omega + i0^+}]$ is not corrected, and the dissipative part of the conductivity correction remains the same as given by (S22).

We now proceed to compute the vertex corrections to the conductivity. For the vertex corrections to the conductivity, we note that the $g'$ and $v$ vertices never contribute in the large $N$ limit due to the insertion of their non-momentum conserving lines decoupling the momentum integrals in the two current vertices of the current correlation function. This causes such diagrams to vanish as the velocity factors in the current vertices are odd under inversion of $\mathbf{k} \to -\mathbf{k}$ whereas the fermion propagators are even under the same. Therefore, only the vertex corrections that solely involve $g$ vertices matter in the large $N$ limit. At order $g^2$, there is only one, the Maki-Thompson diagram (Fig. 1c of the main text), given by

$$\frac{1}{N}\text{Re}[\sigma_{V,g}(\omega)] = \frac{\text{Im}[(f_{V,g}(i\omega) - f_{V,g}(0))_{i\omega \to \omega + i0^+}]}{\omega}$$

$$f_{V,g}(i\omega) = -g^2 v_F^2 k_F^2 \int_{-\infty}^{\infty} \frac{d\omega_1}{2\pi} \int_{-\infty}^{\infty} \frac{d\omega_2}{2\pi} \int_{-\pi}^{\pi} \frac{d\theta_1 \cos\theta_1}{2\pi} \int_{-\pi}^{\pi} \frac{d\theta_2 \cos\theta_2}{2\pi} \int_{-\infty}^{\infty} \frac{dk_1}{2\pi} \int_{-\infty}^{\infty} \frac{dk_2}{2\pi} \frac{1}{i\frac{\Gamma}{2}\text{sgn}(\omega_1) - v_F k_1}$$

$$\times \frac{1}{i\frac{\Gamma}{2}\text{sgn}(\omega_1 + \omega) - v_F k_1} \frac{1}{i\frac{\Gamma}{2}\text{sgn}(\omega_2) - v_F k_2} \frac{1}{i\frac{\Gamma}{2}\text{sgn}(\omega_2 + \omega) - v_F k_2}$$

$$\times \frac{1}{(k_F + k_1)^2 + (k_F + k_2)^2 + c_d|\omega_1 - \omega_2| - 2(k_F + k_1)(k_F + k_2)\cos(\theta_1 - \theta_2)}. \quad (S23)$$

The angular integrals can be computed first

$$\int_{-\pi}^{\pi} \frac{d\theta_1 \cos\theta_1}{2\pi} \int_{-\pi}^{\pi} \frac{d\theta_2 \cos\theta_2}{2\pi} \frac{1}{(k_F + k_1)^2 + (k_F + k_2)^2 + c_d|\omega_1 - \omega_2| - 2(k_F + k_1)(k_F + k_2)\cos(\theta_1 - \theta_2)}$$

$$
= \frac{1}{2} \int_{-\pi}^{\pi} \frac{d\theta_1}{2\pi} \frac{\cos\theta_1}{(k_F+k_1)^2 + (k_F+k_2)^2 + c_d|\omega_1-\omega_2| - 2(k_F+k_1)(k_F+k_2)\cos\theta_1}
$$
$$
\simeq \frac{1}{4k_F \sqrt{(k_1-k_2)^2 + c_d|\omega_1-\omega_2|}} + \ldots , \tag{S24}
$$

where the ... corresponds to terms that are not singular as $k_1,\omega_1 \to k_2,\omega_2$, and are also $\mathcal{O}(1/k_F^2)$ and smaller. These terms may therefore be ignored just like they were in the computation of $\Sigma_g(i\omega,\mathbf{k})$. Since only the singular (as $k,\omega \to 0$) boson fluctuations contribute to the marginal Fermi liquid form of $\Sigma_g(i\omega,\mathbf{k})$, these higher order non-singular corrections will not give rise to any $\omega$ or $T$-linear transport scattering rates (instead contributing higher powers of $\omega, T$), and are therefore of no interest to us. Because of the additional suppression of these non-singular corrections by powers of $1/k_F$ both here and in $\Sigma_g(i\omega,\mathbf{k})$, the conductivity corrections that they induce will also not be extensive in the number of electrons in the system. Substituting (S24) into (S23), we can then write

$$
f_{V,g}(i\omega) = -\frac{g^2 v_F^2 k_F}{4} \int_{-\infty}^{\infty} \frac{d\omega_1}{2\pi} \int_{-\infty}^{\infty} \frac{d\omega_2}{2\pi} \int_{-\infty}^{\infty} \frac{dk_1}{2\pi} \int_{-\infty}^{\infty} \frac{dk_2}{2\pi} \frac{1}{i\frac{\Gamma}{2}\mathrm{sgn}(\omega_1+\omega_2) - v_F(k_1+k_2)}
$$
$$
\times \frac{1}{i\frac{\Gamma}{2}\mathrm{sgn}(\omega_1+\omega_2+\omega) - v_F(k_1+k_2)} \frac{1}{i\frac{\Gamma}{2}\mathrm{sgn}(\omega_2) - v_F k_2} \frac{1}{i\frac{\Gamma}{2}\mathrm{sgn}(\omega_2+\omega) - v_F k_2} \frac{1}{\sqrt{k_1^2 + c_d|\omega_1|}}. \tag{S25}
$$

We now integrate over $k_2$: this results in a complicated function of $\omega,\omega_1,\omega_2$ that switches discontinuously between different values depending on the signs of various linear combinations of $\omega,\omega_1,\omega_2$. Nevertheless, it is then possible to proceed by an additional step and and also integrate over $\omega_2$. This gives

$$
\int_{-\infty}^{\infty} \frac{d\omega_2}{2\pi} \int_{-\infty}^{\infty} \frac{dk_2}{2\pi} \frac{1}{i\frac{\Gamma}{2}\mathrm{sgn}(\omega_1+\omega_2) - v_F(k_1+k_2)} \frac{1}{i\frac{\Gamma}{2}\mathrm{sgn}(\omega_1+\omega_2+\omega) - v_F(k_1+k_2)}
$$
$$
\times \frac{1}{i\frac{\Gamma}{2}\mathrm{sgn}(\omega_2) - v_F k_2} \frac{1}{i\frac{\Gamma}{2}\mathrm{sgn}(\omega_2+\omega) - v_F k_2}
$$
$$
= \Theta(|\omega_1|-|\omega|)\frac{i\Gamma\mathrm{sgn}(\omega_1)(|\omega_1|-2|\omega|) + v_F k_1|\omega|}{2\pi v_F \Gamma(v_F k_1 - i\mathrm{sgn}(\omega_1)\Gamma)^3} + \Theta(|\omega|-|\omega_1|)\frac{i\Gamma\mathrm{sgn}(\omega_1)(2|\omega_1|-|\omega|) + v_F k_1|\omega|}{2\pi v_F \Gamma(v_F k_1 - i\mathrm{sgn}(\omega_1)\Gamma)^2(v_F k_1 + i\mathrm{sgn}(\omega_1)\Gamma)}, \tag{S26}
$$

where $\Theta(...)$ is the Heaviside step function. The first (high frequency) term of the last line of (S26) is not interesting. Inserting it into (S25), performing the $k_1$ integral, and then the $\omega_1$ integral with the physical UV frequency cutoff of the $z=2$ theory, $\omega_{\mathrm{UV}} \sim \Gamma^2/(v_F^2 c_d)$, yields a contribution

$$
\frac{1}{\mathcal{N}}\mathrm{Re}[\sigma_{V,g}(\omega)] = -\frac{\mathcal{N} g^2 c_1}{c_d \Gamma}, \tag{S27}
$$

where $c_1 \sim \ln(\omega_{\mathrm{UV}} v_F^2 c_d/\Gamma^2)$ is a positive $\mathcal{O}(1)$ number, whose precise value depends on the numerical constant of proportionality between the physical UV frequency cutoff $\omega_{\mathrm{UV}}$ and $\Gamma^2/(v_F^2 c_d)$. This contribution is not extensive in the number of fermions in the system (*i.e.* not proportional to $k_F^2$ or $v_F^2$), and is thus only a small (suppressed by $1/k_F^2$) $\omega$ and $T$-independent correction to the static transport scattering rate $\Gamma$. We therefore ignore it.

The second term of the last line of (S26) is important. Inserting it into (S25) and performing the $k_1$ and $\omega_1$ integrals yields the contribution

$$
f_{V,g}(i\omega) - f_{V,g}(0) = -\frac{g^2 v_F k_F}{2\pi \Gamma} \int_0^{|\omega|} \frac{d\omega_1}{2\pi} \int_{-\infty}^{\infty} \frac{dk_1}{2\pi} \left( \frac{\Gamma^2(|\omega|-2\omega_1) + v_F^2 k_1^2|\omega|}{(\Gamma^2 + v_F^2 k_1^2)^2} + \frac{2\Gamma^2 \omega_1}{(\Gamma^2 + v_F^2 k_1^2)^2} \right) \frac{1}{\sqrt{k_1^2 + c_d|\omega_1|}}
$$
$$
= -\frac{g^2 v_F k_F}{2\pi^2 \Gamma^2}|\omega| \int_0^{|\omega|} \frac{d\omega_1}{2\pi} \frac{\sec^{-1}\left(\frac{v_F \sqrt{c_d|\omega_1|}}{\Gamma}\right)}{\sqrt{v_F^2 c_d|\omega_1| - \Gamma^2}}
$$

$$
\begin{aligned}
&= -\frac{\mathcal{N}g^2}{2\pi^2 c_d \Gamma^2}|\omega|\left(\Gamma \ln\left(\frac{4\Gamma^2}{c_d v_F^2 |\omega|}\right) + 2i\sqrt{\Gamma^2 - c_d v_F^2 |\omega|}\sec^{-1}\left(\frac{v_F\sqrt{c_d|\omega|}}{\Gamma}\right)\right) \\
&= -\frac{\mathcal{N}v_F^2 g^2}{8\pi^2 \Gamma^3}\omega^2 \ln\left(\frac{16e^2 \Gamma^4}{v_F^4 c_d^2 \omega^2}\right), \quad |\omega| \to 0; \\
\frac{1}{N}\mathrm{Re}[\sigma_{V,g}(\omega)] &= \frac{\mathcal{N}v_F^2 g^2 |\omega|}{8\pi \Gamma^3}.
\end{aligned} \tag{S28}
$$

This contribution exactly cancels (S22). This is expected since the singular low-momentum boson fluctuations just lead to forward scattering that does not relax current.

Therefore, perturbatively, we have

$$
\begin{aligned}
N\mathrm{Re}\left[\frac{1}{\sigma(\omega)}\right] &= N\mathrm{Re}\left[\frac{1}{\sigma_v(\omega) + \sigma_{\Sigma,g'}(\omega)}\right] \\
&= N\mathrm{Re}\left[\frac{1}{\sigma_v(\omega)\left(1 + \frac{\sigma_{\Sigma,g'}(\omega)}{\sigma_v(\omega)}\right)}\right] \simeq N\mathrm{Re}\left[\frac{1}{\sigma_v(\omega)}\right] - N\mathrm{Re}\left[\frac{\sigma_{\Sigma,g'}(\omega)}{\sigma_v^2(\omega)}\right], \\
&\simeq \frac{1}{\mathcal{N}v_F^2}\left[2\Gamma + \frac{g'^2 |\omega|}{4}\right],
\end{aligned} \tag{S29}
$$

where we used $\sigma_v(\omega) = \mathcal{N}v_F^2/(2\Gamma - 2i\omega)$, and retained only the leading frequency-dependent corrections in the final expression. This gives the much sought after linear-in-energy correction to the static impurity scattering rate $\Gamma$.

When $T > 0$, the $z=2$ boson gains a thermal mass $m_b^2(T) \sim c_d T \ln\ln(\Gamma^2/(v_F^2 c_d T))/\ln(\Gamma^2/(v_F^2 c_d T))$ [1, 3, 4], with $D(i\omega, \mathbf{q}) = 1/(q^2 + c_d|\omega| + m_b^2(T))$. Because $m_b^2(T)$ is not $\gg T \sim |\omega| \sim q^2/c_d$ at low $T$, the low-frequency and low-momentum boson fluctuations do not become any less singular when $T > 0$. Thus, the singular boson fluctuations continue to induce only forward scattering of the fermions, whose contributions to the conductivity continue to cancel between $\sigma_{\Sigma,g}$ and $\sigma_{V,g}$, just like as demonstrated above. The finite temperature conductivity and transport scattering rate are then also simply computed using $\Sigma_{g'}$, which was also done in Refs. [1, 3]. This gives a $\sim \mathcal{N}g'^2 T \ln\ln(\Gamma^2/(v_F^2 c_d T))$ correction to the static impurity transport scattering rate $\Gamma$ [1, 3].

While the $T$-linear correction to the transport scattering rate arises only from $\Sigma_{g'}$, both $\Sigma_g$ and $\Sigma_{g'}$ contribute to the effective mass renormalization of the fermions. Since these are both of marginal Fermi liquid form, the effective mass renormalization is $m^*/m \sim (a_1 g^2/\Gamma + a_2 \mathcal{N}g'^2)\ln(\Gamma^2/(v_F^2 c_d T))$, where the numbers $a_{1,2} \sim \mathcal{O}(1)$ [1]. Then, following the steps in Sec. VIII D of Ref. [1], we obtain the result for the constant of proportionality $\alpha$ between $1/\tau_{\mathrm{tr}}^*$ and $k_B T/\hbar$ given in the main text (Eq. (13)).

Considering diagrams with four interaction vertices, we find the two Aslamazov-Larkin diagrams (Fig. 1d,e of the main text) in the large $N$ limit (which are also encountered in the kernel of the full ladder resummation described in Section IV of Ref. [2]). We will show below that the sum of these two diagrams does not correct the conductivity when $k_F$ is large. The other diagrams are the two rung $g$ ladder, and additional insertions of the $g$ and $g'$ self energies and ladders. From the computation of Section IV of Ref. [2], it can be established that the resummation of such non-Aslamazov-Larkin terms essentially just renormalizes the current-relaxing electron scattering rate, while canceling the current-conserving forward scattering at all orders. In particular, since $g'$ does not generate new vertex corrections [5]. Therefore, the renormalization of the transport scattering rate described by (S29) actually holds to all orders in perturbation theory in the large $N$ limit.

We now demonstrate the nullification of the sum of the two Aslamazov-Larkin diagrams in Fig. 1d,e of the main text in the large Fermi energy or large $k_F$ limit. We can express the sum of the two order $g^4$ Aslamazov-Larkin

diagrams in this limit as

$$f_{\text{AL},g}(i\omega) = g^4 v_F^2 k_F^2 \int_{-\infty}^{\infty} \frac{d\omega_1}{2\pi} \int_{-\infty}^{\infty} \frac{d\omega_2}{2\pi} \int_{-\infty}^{\infty} \frac{d\omega_3}{2\pi} \int_{-\pi}^{\pi} \frac{d\theta_1 \cos\theta_1}{2\pi} \int_{-\pi}^{\pi} \frac{d\theta_2}{2\pi} \int_{-\pi}^{\pi} \frac{d\theta_3 \cos\theta_3}{2\pi} \int_{-\infty}^{\infty} \frac{dk_1}{2\pi} \int_{-\infty}^{\infty} \frac{k_2 dk_2}{2\pi} \int_{-\infty}^{\infty} \frac{dk_3}{2\pi}$$
$$\frac{1}{k_2^2 + c_d|\omega_2 + \frac{\omega}{2}|} \frac{1}{k_2^2 + c_d|\omega_2 - \frac{\omega}{2}|} \frac{1}{i\frac{\Gamma}{2}\text{sgn}\left(\omega_1 + \frac{\omega}{2}\right) - v_F k_1} \frac{1}{i\frac{\Gamma}{2}\text{sgn}\left(\omega_1 - \frac{\omega}{2}\right) - v_F k_1}$$
$$\times \frac{1}{i\frac{\Gamma}{2}\text{sgn}\left(\omega_3 + \frac{\omega}{2}\right) - v_F k_3} \frac{1}{i\frac{\Gamma}{2}\text{sgn}\left(\omega_3 - \frac{\omega}{2}\right) - v_F k_3} \frac{1}{i\frac{\Gamma}{2}\text{sgn}(\omega_1 - \omega_2) - v_F k_1 + v_F k_2 \cos(\theta_1 - \theta_2)}$$
$$\times \left( \frac{1}{i\frac{\Gamma}{2}\text{sgn}(\omega_3 + \omega_2) - v_F k_3 - v_F k_2 \cos(\theta_3 - \theta_2)} + \frac{1}{i\frac{\Gamma}{2}\text{sgn}(\omega_3 - \omega_2) - v_F k_3 + v_F k_2 \cos(\theta_3 - \theta_2)} \right). \tag{S30}$$

We can then see that the quantity in brackets on the last line is odd under $k_3, \omega_3 \to -k_3, -\omega_3$, whereas all the other terms multiplying it are even under the same, which renders the whole integrand odd under $k_3, \omega_3 \to -k_3, -\omega_3$. Therefore the integral over $k_3, \omega_3$ (and hence $f_{\text{AL},g}(i\omega)$) vanishes identically. In fact, this continues to occur when the self-energies $\Sigma_g, \Sigma_{g'}$ and the bare $i\omega$ term are included in the fermion propagators, as these are all odd in the Matsubara frequency. The cancellation of the Aslamazov-Larkin diagrams is therefore completely self-consistent in the large $k_F$ limit. The only non-zero contributions to the sum of the Aslamazov-Larkin diagrams comes from going away from the large $k_F$ limit, by including corrections to the fermion current vertex factors and fermion dispersions. The resulting corrections to the conductivity are therefore not extensive in the number of fermions in the system, and are therefore not important to us. A computation of these non-extensive contributions is nevertheless carried out in Section IV.C.3 of Ref. [2], where it is demonstrated that they indeed lead to a small correction to the transport scattering rate that is $\mathcal{O}(\mathcal{N} g^4 \omega^2/(\Gamma^2 k_F^2))$.

Finally, we consider the case of $v = 0$ but $g, g' \neq 0$. In this case, the non-interacting conductivity (Fig. 1a of the main text) is trivially given by

$$\frac{1}{\mathcal{N}} \sigma_0(i\omega) = \frac{\mathcal{N} v_F^2}{2\omega}. \tag{S31}$$

The contributions of the momentum-independent $\Sigma_g(i\omega, \mathbf{k}) \simeq -ic_f \text{sgn}(\omega)|\omega|^{2/3}$ and $\Sigma_{g'}(i\omega)$ (S19) to the conductivity within perturbation theory (i.e. Fig. 1b of the main text) are also straighforwardly computed as in (S21), and are given by

$$\frac{1}{\mathcal{N}} \sigma_{\Sigma,g'}(i\omega) = -\frac{\mathcal{N}^2 v_F^2 g'^2}{24\pi\omega} \ln\left(\frac{e^3 \tilde{\Lambda}^6}{c_b^2 \omega^2}\right),$$
$$\frac{1}{\mathcal{N}} \sigma_{\Sigma,g}(i\omega) \simeq -\frac{3\mathcal{N} v_F^2 c_F}{5\omega|\omega|^{1/3}} = -\frac{3\mathcal{N} v_F g^2}{10\pi\sqrt{3} c_b^{1/3} \omega|\omega|^{1/3}}.$$

As before, only vertex corrections with $g$ vertices contribute in the large $N$ limit. The contributions to $\sigma_{V,g}(i\omega)$ from the most singular boson fluctuations can be computed using the theory of antipodal patches described in Ref. [1]. We then have

$$\frac{1}{\mathcal{N}} \sigma_{V,g}(i\omega) = \frac{v_F^2 g^2}{\omega} \sum_{s=\pm} \int_{-\infty}^{\infty} \frac{d\omega_1}{2\pi} \int_{-\infty}^{\infty} \frac{d\omega_2}{2\pi} \int_{-\infty}^{\infty} \frac{dk_{1x}}{2\pi} \int_{-\infty}^{\infty} \frac{dk_{2x}}{2\pi} \int_{-\infty}^{\infty} \frac{dk_{1y}}{2\pi} \int_{-\infty}^{\infty} \frac{dk_{2y}}{2\pi} \frac{|k_{1y} - k_{2y}|}{|k_{1y} - k_{2y}|^3 + c_b|\omega_1 - \omega_2|}$$
$$\times \frac{1}{i\omega_1 - sv_F k_{1x} - \kappa \frac{k_{1y}^2}{2}} \frac{1}{i(\omega_1 + \omega) - sv_F k_{1x} - \kappa \frac{k_{1y}^2}{2}} \frac{1}{i\omega_2 - sv_F k_{2x} - \kappa \frac{k_{2y}^2}{2}} \frac{1}{i(\omega_2 + \omega) - sv_F k_{2x} - \kappa \frac{k_{2y}^2}{2}}$$
$$= \frac{2g^2}{\omega} \int_{-\infty}^{\infty} \frac{d\omega_1}{2\pi} \int_{-\infty}^{\infty} \frac{d\omega_2}{2\pi} \left( \frac{\text{sgn}(\omega_1 + \omega) - \text{sgn}(\omega_1)}{2\omega} \right) \left( \frac{\text{sgn}(\omega_2 + \omega) - \text{sgn}(\omega_2)}{2\omega} \right) \int_{-\infty}^{\infty} \frac{dk_{1y}}{2\pi} \int_{-\infty}^{\infty} \frac{dk_{2y}}{2\pi}$$

$$\times \frac{|k_{1y} - k_{2y}|}{|k_{1y} - k_{2y}|^3 + c_b|\omega_1 - \omega_2|}$$

$$= \frac{2g^2 \Lambda_y}{\omega} \int_{-\infty}^{\infty} \frac{d\omega_1}{2\pi} \int_{-\infty}^{\infty} \frac{d\omega_2}{2\pi} \left( \frac{\text{sgn}(\omega_1 + \omega) - \text{sgn}(\omega_1)}{2\omega} \right) \left( \frac{\text{sgn}(\omega_2 + \omega) - \text{sgn}(\omega_2)}{2\omega} \right) \int_{-\infty}^{\infty} \frac{dk_{1y}}{2\pi} \frac{|k_{1y}|}{|k_{1y}|^3 + c_b|\omega_1 - \omega_2|}$$

$$= \frac{3\Lambda_y g^2}{5\sqrt{3}\pi^2 c_b^{1/3} \omega |\omega|^{1/3}} = \frac{3\mathcal{N} v_F g^2}{10\sqrt{3}\pi c_b^{1/3} \omega |\omega|^{1/3}}, \quad \Lambda_y = \int_{-\infty}^{\infty} \frac{dk_{2y}}{(2\pi)} = \frac{\pi \mathcal{N} v_F}{2}, \tag{S32}$$

where we shifted $k_{1y} \to k_{1y} + k_{2y}$, and fixed the cutoff on the patch size $\Lambda_y$ so that the correct non-interacting conductivity (S31) is obtained from the theory of antipodal patches when $v = g = g' = 0$. Therefore, we find that the most singular contribution in $\sigma_{V,g}(i\omega)$ cancels with $\sigma_{\Sigma,g}(i\omega)$. Additionally, it can also be shown that within the theory of antipodal patches, where $\sigma_{V,g}(i\omega)$ is restricted to its most singular contribution, the sum of the two Aslamazov-Larkin diagrams vanishes due to an odd/even cancellation of integrands like in (S30). Therefore, we obtain the results in Eq. (11) of the main text.

The above conclusions are however even stronger, because the Prange-Kadanoff reduction of Section III.E.1 of Ref. [2] is still valid when $v = 0, g \neq 0, g' \neq 0$. It then follows from the analysis in Section III.E.1 of Ref. [2] that the less singular contributions in $\sigma_{V,g}(i\omega)$ arising from going beyond the theory of antipodal patches cancel with the Aslamazov-Larkin diagrams exactly like they did when $v = 0, g \neq 0, g' = 0$ in Section III.E.1 of Ref. [2] [6]. Therefore, the antipodal patch theory essentially reproduces the results from the full theory going beyond antipodal patches for the low-frequency behavior of the optical conductivity in the $v = 0$ case, even though it doesn't capture all the physics of the system correctly. As was the case for $v \neq 0, g \neq 0, g' \neq 0$, the perturbative results for the conductivity here are also valid to all orders in perturbation theory in the large $N$ limit, for the same reasons as before.

[1] I. Esterlis, H. Guo, A. A. Patel, and S. Sachdev, *Large N theory of critical Fermi surfaces*, Phys. Rev. B **103**, 235129 (2021), arXiv:2103.08615 [cond-mat.str-el].

[2] H. Guo, A. A. Patel, I. Esterlis, and S. Sachdev, *Large-N theory of critical Fermi surfaces. II. Conductivity*, Phys. Rev. B **106**, 115151 (2022), arXiv:2207.08841 [cond-mat.str-el].

[3] E. E. Aldape, T. Cookmeyer, A. A. Patel, and E. Altman, *Solvable theory of a strange metal at the breakdown of a heavy Fermi liquid*, Phys. Rev. B **105**, 235111 (2022), arXiv:2012.00763 [cond-mat.str-el].

[4] A. A. Patel and S. Sachdev, *DC resistivity at the onset of spin density wave order in two-dimensional metals*, Phys. Rev. B **90**, 165146 (2014), arXiv:1408.6549 [cond-mat.str-el].

[5] The only net effect of $g'$ in (4.20) of Ref. [2] is to add a term analogous to (4.23) of Ref. [2], but involving $\Sigma_{g'}$ instead of $\Sigma_g$, to $W_{\Sigma,FF}^{-1}$. Analogous to $W_{\text{dis}}[F]$, terms like $W_{g'}[F]$ involve only the zeroth angular harmonic, and can therefore be ignored, as the current vertex $\Gamma^x$ is in the first harmonic sector. This is completely equivalent to $g'$ not generating any new vertex corrections in the perturbative computation.

[6] Because the $|\omega|$ damping term induced by $g'$ in the boson propagator is sub-leading to the $|\omega|/|\mathbf{q}|$ damping term induced by $g$, it can be neglected as we have always done, and the computations analogous to Section III.E.1 of Ref. [2] are then actually identical to those in Section III.E.1 of Ref. [2] even in this case of $v = 0, g \neq 0, g' \neq 0$.


\fi

\end{document}